\documentclass{aa}  
\usepackage{natbib}
\usepackage{booktabs}
\usepackage{blindtext}
\usepackage{hyperref}
\usepackage{booktabs}
\usepackage{multirow}
\usepackage{tabularx}
\usepackage{adjustbox}
\usepackage[utf8]{inputenc}
\usepackage{graphicx}
\usepackage{txfonts}

\bibpunct{(}{)}{;}{a}{}{,}

\begin{document}

   \title{How plasma coupling and convective-zone depth shape the rotation of solar-mass stars}

   \author{A. Brito
          \inst{1,2,3} 
          \and
          I. Lopes \inst{2}
          }
          
    \institute{Instituto Superior de Gest\~ao\\ Rua Prof. Reinaldo dos Santos 46 A, 1500-552, Lisboa, Portugal\\
 	              \email{anabrito@isg.pt}
 	         \and
 	              Centro de Astrof\'{\i}sica e Gravita\c c\~ao  - CENTRA, Departamento de F\'{\i}sica, Instituto Superior T\'ecnico \\ IST, Universidade de Lisboa - UL, Av. Rovisco Pais 1, 1049-001 Lisboa, Portugal\\
 	              \email{ilidio.lopes@tenico.ulisboa.pt}
 	          \and
 	              Atlântica -- Instituto Universitário, 2730-036, Barcarena, Portugal
 	             }         

   \date{}

 \abstract
{Stellar rotation on the main sequence is a complex function of mass and age, displaying multiple regimes whose physical origin remains only partially understood. In particular, the connection between the diversity of observed rotation rates and the internal structure and thermodynamic properties of stellar interiors is still unclear.}
{We investigated how the depth of the convective zones and the degree of plasma coupling, quantified through the plasma coupling parameter, relate to the observed rotation rates of solar-mass stars.}
{We used a grid of $1 \, M_\odot$ MESA stellar models with a wide range of metallicities to identify the best-matching models for 243 main-sequence stars with measured rotation periods. We then examined correlations between their rotation rates and both the structural properties of the convective zones and the corresponding convective plasma coupling parameter.
}
{
For this sample, rotation rates show only weak correlations with either the convective-zone depth or the plasma coupling parameter when considered independently. However, during the first two-thirds of the main-sequence lifetime, the correlation strengthens when both factors are considered jointly through a combined convective coupling index, indicating a moderate and statistically significant relationship. For older stars, these correlations weaken and lose significance, although the thermodynamic component becomes relatively more influential. These trends suggest that microphysical plasma properties may contribute to the regulation of angular momentum loss and may be connected to the onset of weakened magnetic braking.

}
{}

   \keywords{Stars: evolution --
                Stars: interiors --
                Stars: low-mass -- Stars: rotation -- Stars: solar-type -- Stars: activity
               }
	\titlerunning{Plasma coupling, CZ depth, and stellar rotation}
	\authorrunning{Ana Brito \& Ilídio Lopes}
 	\maketitle
 	\nolinenumbers
  % \titlerunning{small title}
%
%________________________________________________________________

\section{Introduction}\label{sec:1}

Low-mass stars with outer convective zones begin their main-sequence evolution rotating rapidly, and as they age, they lose angular momentum through a process known as magnetic braking. In low-mass stars, magnetic braking occurs when dynamo-generated magnetic fields in the outer convective envelope drive magnetized stellar winds that extract angular momentum over the course of main-sequence evolution \citep{1967ApJ...150..551K, 1967ApJ...148..217W, 1972ApJ...171..565S, 1988ApJ...333..236K}. Rotation in these stars also exhibits a clear relationship with several magnetic activity indicators, such as nonthermal emission, X--ray emission, and rotational modulation \citep[e.g.,][]{1984ApJ...279..763N, 2003A&A...397..147P, 2021ApJS..255...17S}. Slower rotation is generally associated with weaker magnetic activity, in what is commonly termed the rotation–activity relationship. This relationship comprises two regimes: the saturated regime, where magnetic activity is independent of rotation, and the unsaturated regime, where magnetic activity decreases with increasing rotation period \citep[e.g.,][]{2011ApJ...743...48W, 2025A&A...699A.251Y}. The transition between these regimes occurs at a critical value of the stellar Rossby number, $\text{Ro} = \frac{P_{\text{rot}}}{\tau_c}$, where $P_{\mathrm{rot}}$ is the rotation period and $\tau_c$ is the convective turnover time \citep[e.g.,][]{2024A&A...684A.156N}.

Solar-like stars typically enter the main sequence in the saturated regime, characterized by rapid rotation and intense magnetic activity. As they evolve, magnetized stellar winds extract angular momentum, leading the stars to transition into the unsaturated regime. This loss of angular momentum, and the resulting spin-down, occurs more rapidly in fast rotators than in slower ones. The spin-down rate also depends on stellar mass, with lower-mass stars taking longer to converge onto the so-called slow-rotation sequence \citep{2013A&A...556A..36G, 2015A&A...577A..98G}. Once stars reach this sequence, their rotation periods increase approximately as the square root of stellar age: $P_{\text{rot}} \propto \sqrt{t}$ \citep{1972ApJ...171..565S}. These observational trends form the basis of the technique known as gyrochronology, which uses stellar rotation as a proxy for age \citep{2003ApJ...586..464B, 2010ApJ...721..675B, 2008ApJ...687.1264M, 2007ApJ...669.1167B, 2014ApJ...790L..23D, 2021A&A...652A..60F, 2024AJ....167..159L}. This method is particularly valuable for main-sequence stars for which asteroseismic age determinations are not available. 

The launch of the Kepler space telescope provided a wealth of high-precision asteroseismic measurements \citep{2014ApJS..211...24M, 2015MNRAS.450.1787A}. Asteroseismology is a powerful technique for probing the internal structure and evolutionary state of stars through their oscillation mode frequencies \citep[e.g.,][]{2010aste.book.....A, 2019MNRAS.488.1558B}. The high-quality oscillation data from Kepler allow us to measure stellar rotation periods and to construct calibrated theoretical stellar models constrained by seismic observables \citep[e.g.,][]{2015A&A...582A..10N, 2015MNRAS.446.2959D}. Comparisons of stellar ages and rotation rates derived from asteroseismology show that older main-sequence stars deviate from the Skumanich law: they rotate faster than predicted by gyrochronology. This phenomenon, known as weakened magnetic braking, indicates that stars beyond roughly half of their main-sequence lifetime no longer obey the standard gyrochronology relations \citep{2016Natur.529..181V, 2016ApJ...826L...2M, 2021NatAs...5..707H}.

It is well established that rotation on the main sequence is a complex function of both mass and age. Moreover, stellar rotation and magnetic activity are tightly interconnected through the activity–rotation relationship. Magnetic activity encompasses a variety of energetic phenomena observed in stellar atmospheres, such as starspots, flares, and coronal mass ejections. Similar to the Sun’s 11-year activity cycle, many other stars exhibit cyclic variations in magnetic activity, with periods ranging from a few years to more than a decade \citep{2018ApJ...855...75R, 2017MNRAS.464.3281W, 2023SSRv..219...54J, 2024A&A...686A..90Z}. The magnetized stellar winds responsible for magnetic braking are themselves a continuous manifestation of this activity. The underlying engine that drives all these phenomena is the stellar dynamo: a physical process operating in stellar interiors that amplifies and sustains magnetic fields \citep[e.g.,][]{2017LRSP...14....4B, 2019ApJ...876...83A, 2023SSRv..219...58K}.

The physical processes driving the evolution of stellar rotation and magnetic activity on the main sequence remain poorly understood, as revealed by discrepancies between observations and theoretical models. In particular, the relationship between the thermodynamic properties of stellar interiors and the observed stellar rotation rates remains largely unexplored. A recent study by \citet{2024A&A...690A.228B} identified a correlation between stellar rotation periods and the thermodynamic properties of the stellar plasma, expressed through the plasma coupling parameter. This dimensionless quantity measures the ratio of the average Coulomb potential energy between particles to their average kinetic (thermal) energy, thereby indicating the degree of coupling within the plasma. In the present work, we aim to further investigate this correlation. Because stellar rotation is a complex function of both mass and age, we restrict our analysis to stars of approximately the same mass, specifically around $1 \, M_\odot$ , to minimize these dependencies. Moreover, since the thickness of the convective zone in such stars is a key factor in the operation of the stellar dynamo \citep[e.g.,][]{2020LRSP...17....4C, 2023SSRv..219...35C}, we explore how a combined index that accounts for both the convective zone extent and the degree of plasma coupling can help explain the observed diversity in stellar rotation periods.

In Section \ref{sec:2}, we describe the selection of stars with observationally determined rotation periods and the input physics adopted for the theoretical models. Section \ref{sec:3} presents the properties of these models, including the diversity in the locations of the bases of the convective zones, stellar ages, and the degree of plasma coupling characterized by the plasma coupling parameter. In this section, we also introduce the convective coupling indicator. We discuss the connection between stellar structure, internal thermodynamics, and rotation rates in Section \ref{sec:4}. Finally, we summarize our conclusions in Section \ref{sec:5}.

\section{Sample selection and theoretical modeling}
\label{sec:2}

\begin{figure}
	\centering
	\includegraphics[width=8cm]{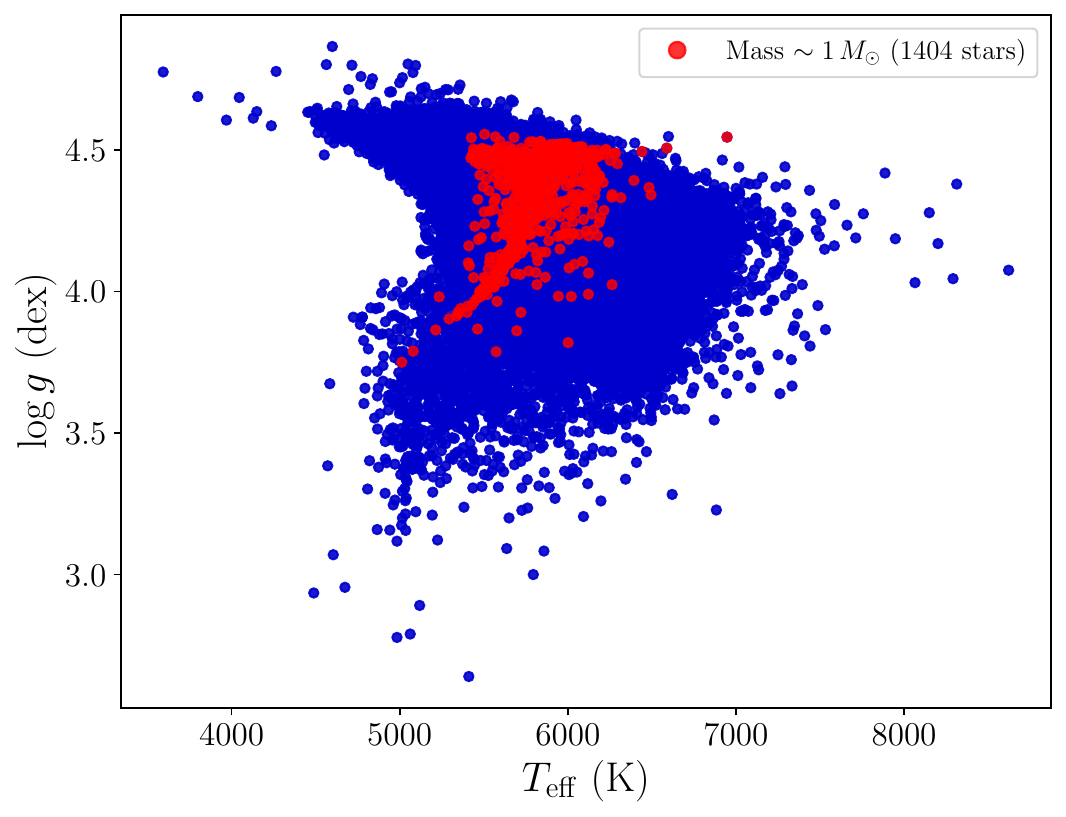}
	\includegraphics[width=8cm]{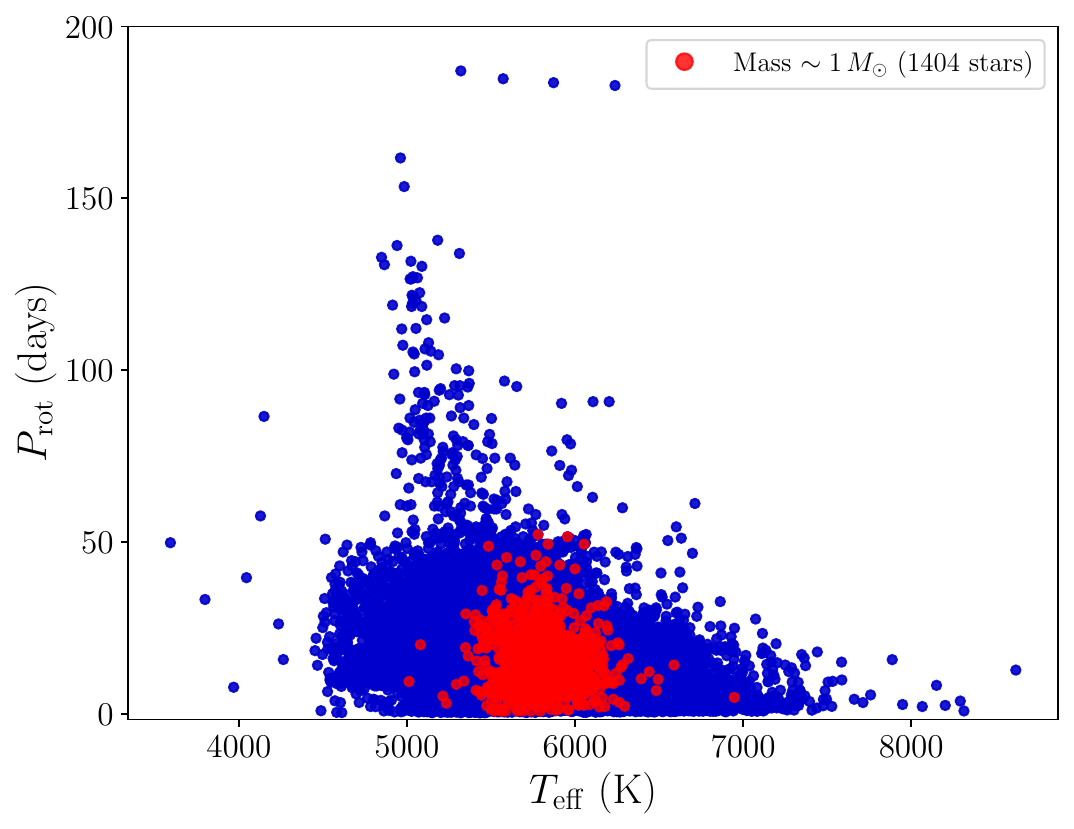}
	\caption{Red dots represent all the stars from the catalog of \citep{2021ApJS..255...17S} with masses in the range  $ 0.99 \, M_\odot< M < 1.005 \,  M_\odot$. }
	\label{fig1}
\end{figure}

The Kepler space telescope enabled the collection of a large sample of stars with measured rotation periods \citep{2021ApJS..255...17S}. The catalog also provides measurements of magnetic activity through the photometric activity index $S_{\text{ph}}$ \citep{2014A&A...562A.124M} for thousands of stars. We combined the observational data from this catalog with the results from our theoretical models to investigate the joint influence of stellar structure and internal thermodynamics on the observed rotation rates. To reduce the complexity of the analysis, we focused on stars with approximately one solar mass. Specifically, we selected only stars with masses in the range $ 0.99 \, M_\odot< M < 1.005 \,  M_\odot$, resulting in a sample of 1404 stars, shown in red in Figure \ref{fig1}.

\begin{figure}
	\centering
	\includegraphics[width=8cm]{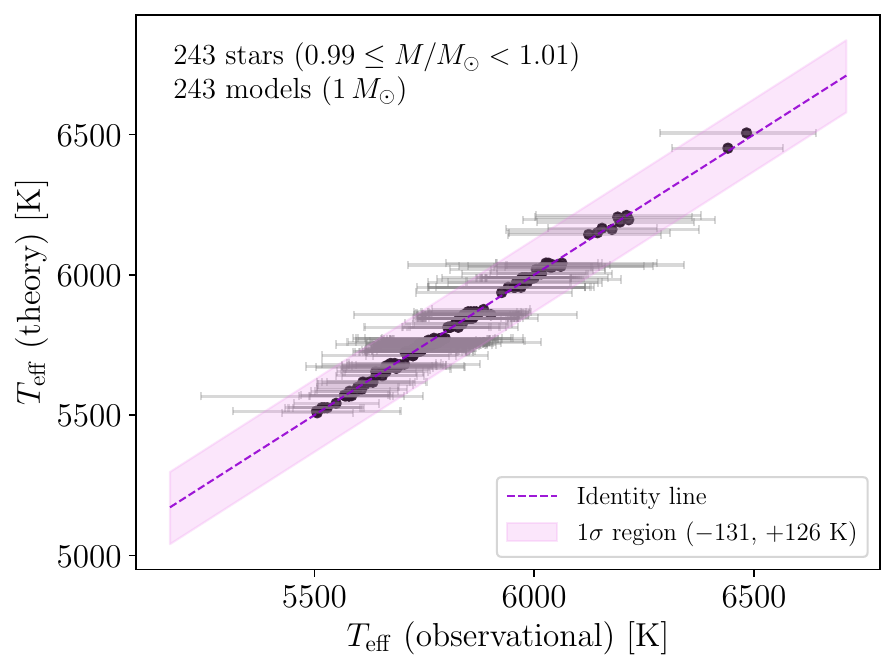}
	\includegraphics[width=8cm]{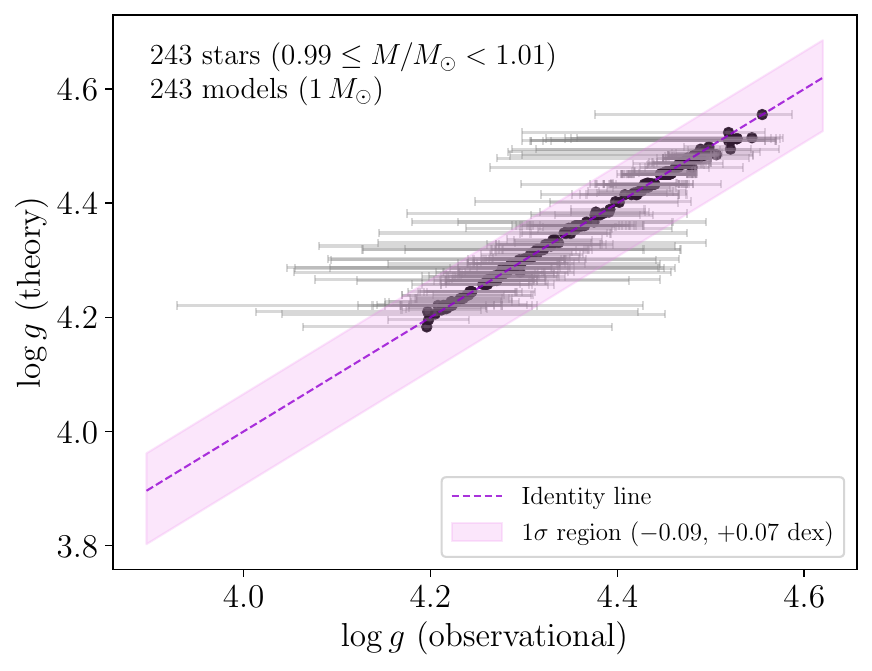}
	\caption{Comparison between observed and best-fit theoretical stellar parameters for our sample of 243 main-sequence stars with masses $ 0.99 \, M_\odot< M < 1.005 \,  M_\odot$, matched to $1 \, M_\odot$ theoretical models. Top panel: Effective temperature (Teff) identity plot. Bottom panel: Surface gravity (log g) identity plot. In both panels, the dashed violet line represents the one-to-one relation, while the shaded violet band indicates the typical $1 \, \sigma$ asymmetric observational uncertainty region. Individual asymmetric error bars for each star are shown in gray. The close clustering of points around the identity lines demonstrates the robustness of our chi-square matching algorithm in recovering physically consistent atmospheric parameters.}
	\label{fig2}
\end{figure}

\subsection{Stellar evolution modeling}

\begin{figure}
	\centering
	\includegraphics[width=8cm]{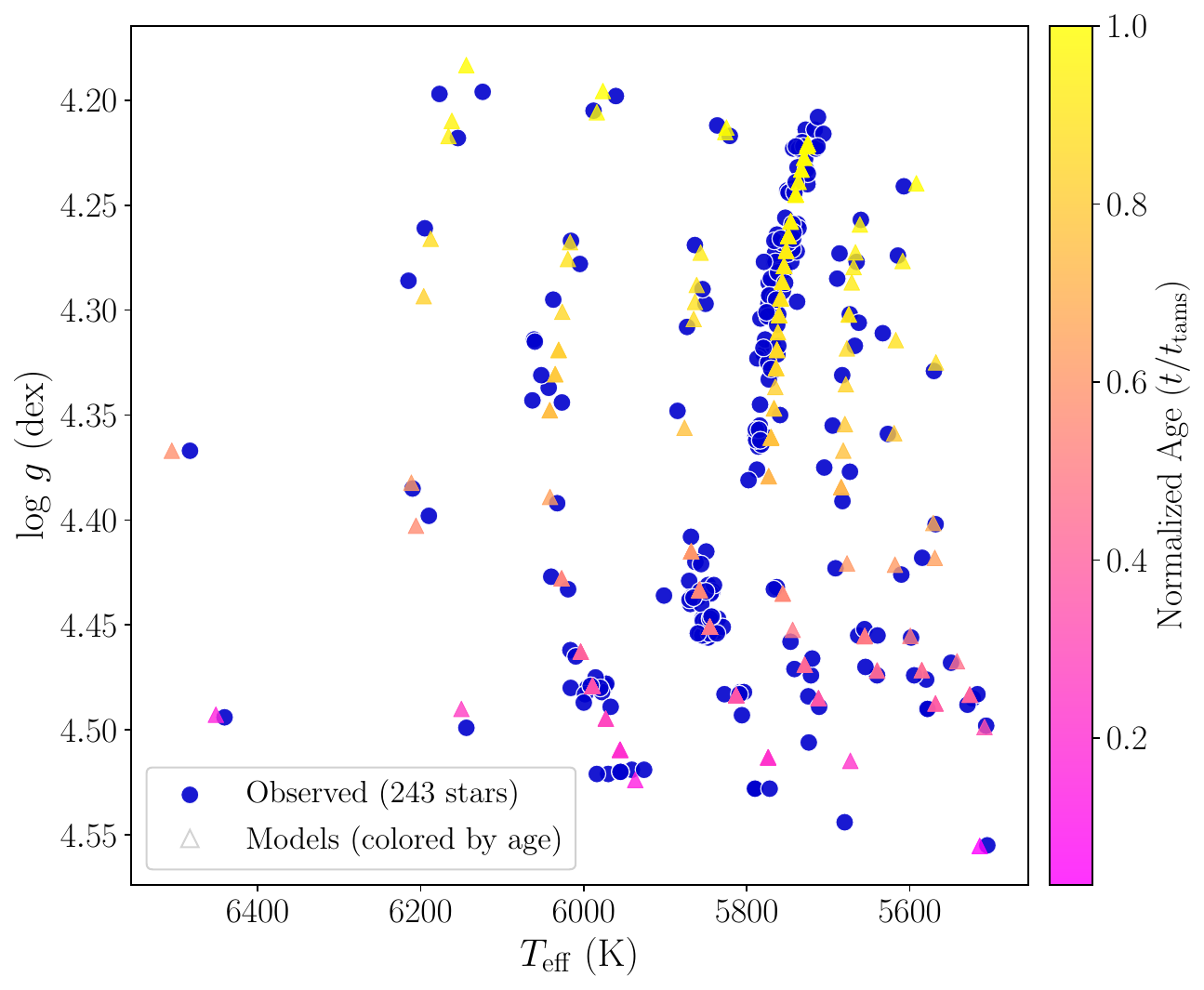}
	\caption{Kiel diagram comparing observed stellar parameters with best-fit theoretical models. The blue circles represent the 243 observed main-sequence stars ($ 0.99 \, M_\odot< M < 1.005 \,  M_\odot$), while the age-colored triangles show their corresponding best-fit $1 \, M_\odot$ theoretical models from our chi-square matching procedure. The color scale indicates the normalized stellar age. The clear progression from younger (pink) to older (yellow) models along the evolutionary track demonstrates that our matching algorithm successfully recovers physically meaningful age sequences.}
	\label{fig3}
\end{figure}

\begin{figure}
	\centering
	\includegraphics[width=8cm]{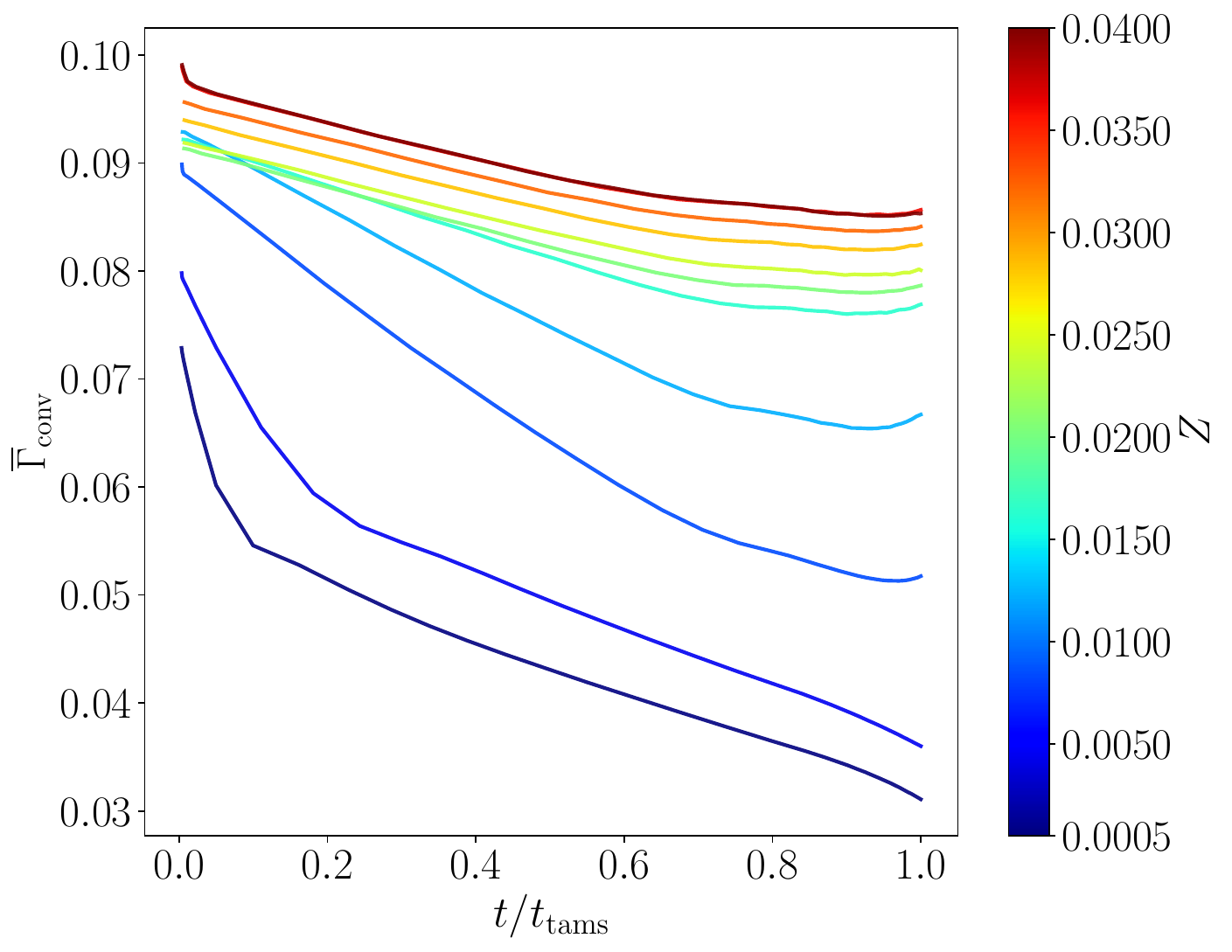}
	\caption{Global convective plasma coupling parameter, $\Gamma_{\mathrm{conv}}$, as a function of normalized main-sequence lifetime ($t/t_{\mathrm{tams}}$) for our set of 243 stellar models of $1 \, M_\odot$ and different metallicities. Each curve corresponds to a distinct metallicity value, as indicated in the legend. Increasing metallicity leads to higher $\Gamma_{\mathrm{conv}}$ values, indicating stronger plasma coupling and more significant collective effects in metal-rich stellar convective zones.}
	\label{fig4}
\end{figure}

\begin{figure}
	\centering
	\includegraphics[width=8cm]{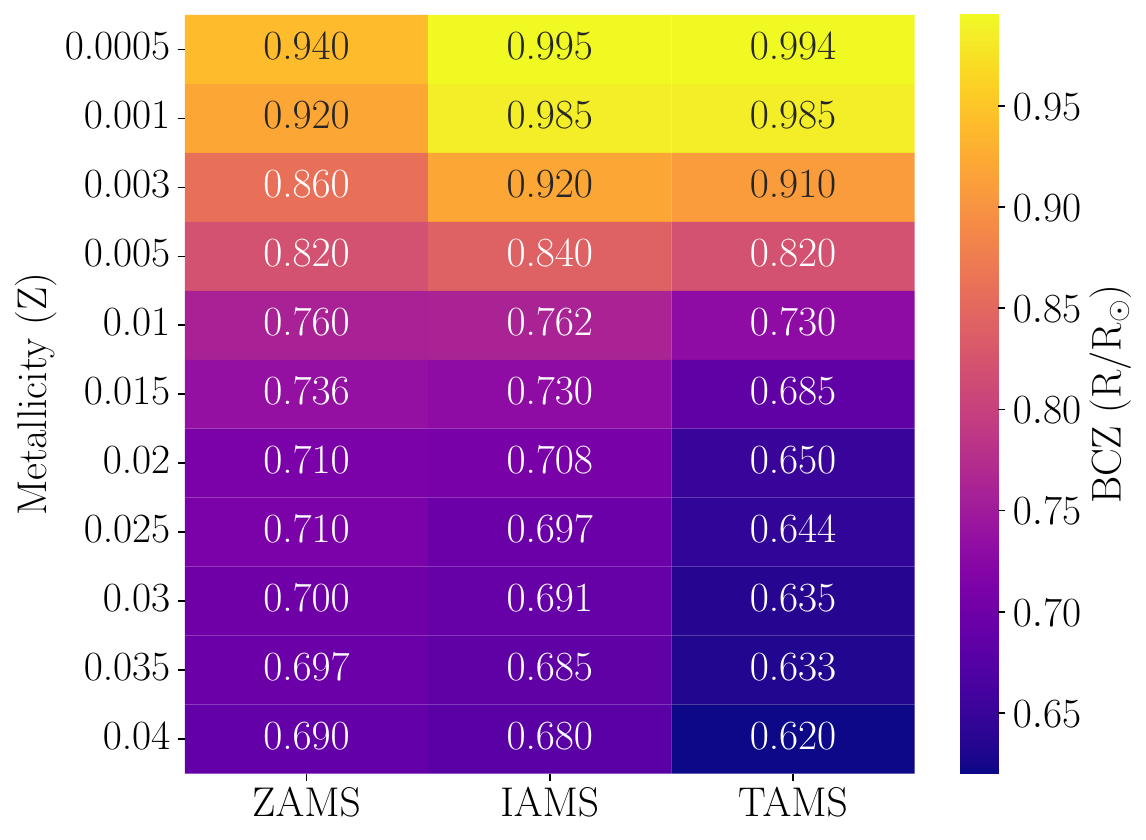}
	\caption{Locations of the base of the convective zone (BCZ), expressed as a fractional stellar radius ($R_{\mathrm{BCZ}}/R_\star$) for $1 \, M_\odot$ stellar models with different metallicities ($Z$). Each line corresponds to a distinct metallicity, with values indicated on the left side. The locations of the BCZ are shown at three evolutionary stages along the main sequence: the zero-age main sequence (ZAMS), intermediate-age main sequence (IAMS), and terminal-age main sequence (TAMS).  The IAMS is defined as the evolutionary stage at which the stellar age corresponds to 50\% of the total main-sequence lifetime ($t/t_{\mathrm{TAMS}} = 0.5$). The figure illustrates the diversity in the depths of the convective zones and how they vary with both age and metallicity. For reference, the MESA standard solar model (computed with $Z=0.02$) predicts $R_{\mathrm{BCZ}}/R_\odot = 0.714$ at the solar age of 4.61 Gyr. }

	\label{fig5}
\end{figure}

\begin{figure}
	\centering
	\includegraphics[width=8cm]{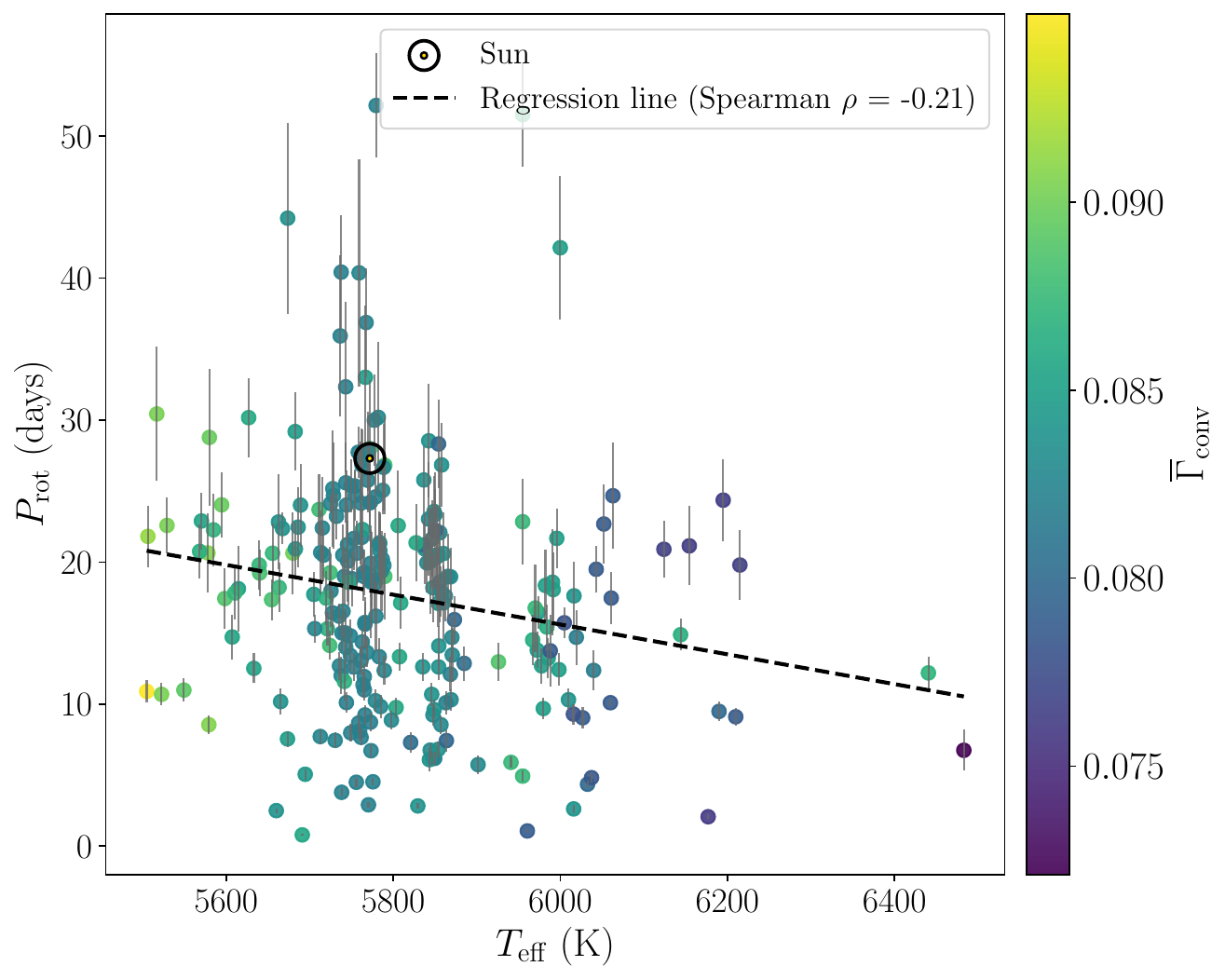}
	\includegraphics[width=8cm]{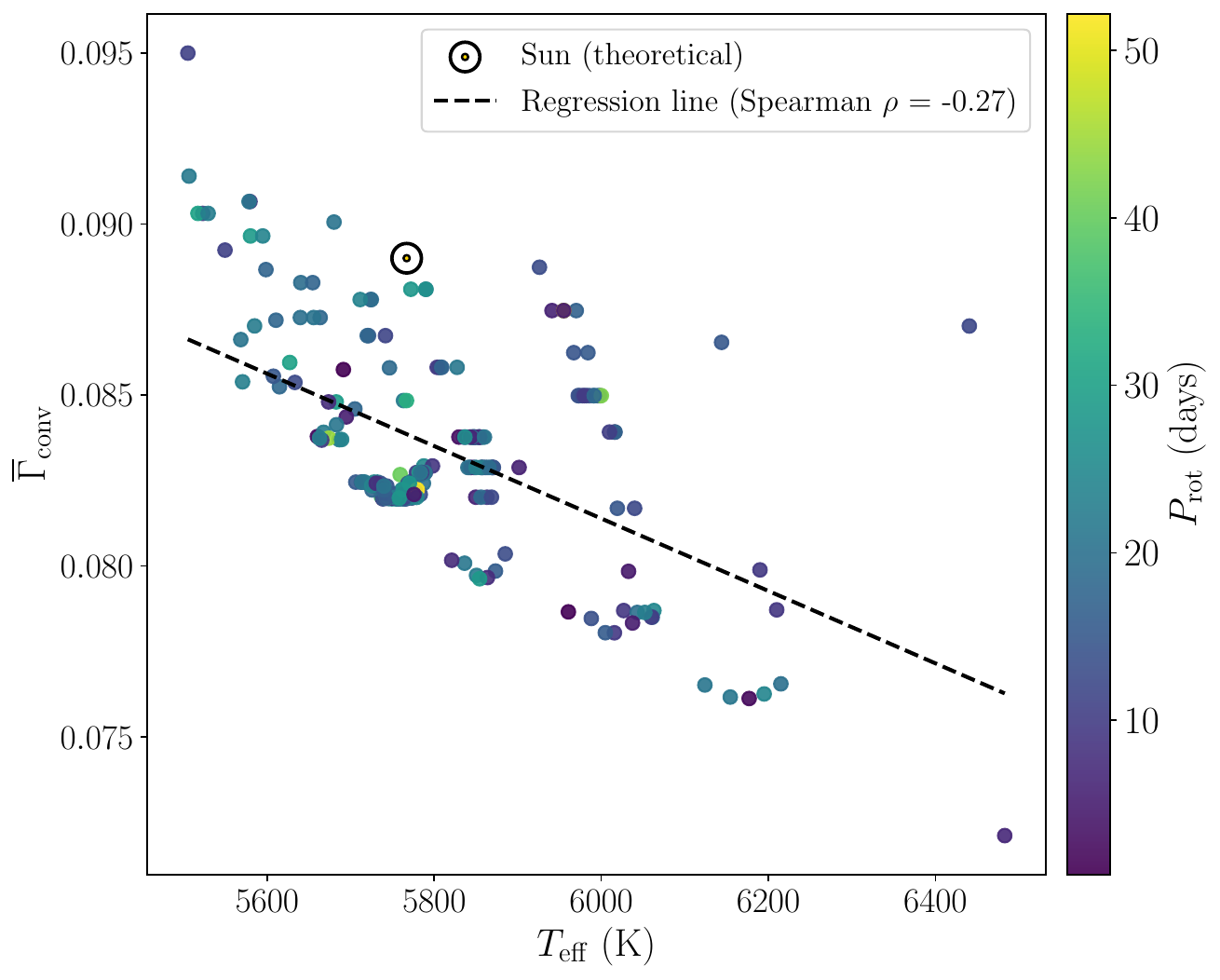}
	\caption{Top panel: Rotation periods as a function of effective temperature for the sample of 243 $1\,M_\odot$ stars described in Section \ref{sec:2}. 
		Data points are color-coded by the values of the convective plasma parameter, and the gray bars indicate the observational uncertainties in the rotation periods. 
		A linear regression fit to the data yields a Spearman correlation coefficient of $\rho = -0.21$ with a p-value of $1.4\times10^{-5}$. 
		Bottom panel: Convective plasma parameter (Eq. \ref{eq4}) as a function of effective temperature for the same set of stars. 
		Data points are color-coded by the measured rotation periods. 
		A linear regression fit yields a Spearman correlation coefficient of $\rho = -0.27$ with a p-value of $1.2\times10^{-3}$. For reference, the solar values are indicated by the symbol $\odot$. The theoretical solar model is computed using the GS98 metallicity.
		}
	\label{fig6}
\end{figure}

To estimate the structural and thermodynamic properties of this sample of stars, we computed a grid of theoretical stellar models with one solar mass using the Modules for Experiments in Stellar Astrophysics (MESA) stellar code, version r24.08.1 \citep{Paxton2011, Paxton2013, Paxton2015, Paxton2018, Paxton2019, 2023ApJS..265...15J}. The metallicities of our models range from $Z=0.0005$ to $Z=0.04$, 
specifically including $Z_i \in \{ 0.0005, 0.001, 0.003, 0.005,$ 
$0.01, 0.015, 0.02, 0.025, 0.03, 0.035, 0.04 \}$. We scaled the initial helium mass fractions with metallicity according to the standard chemical enrichment relation,
\begin{equation}
	Y_i = Y_0 + \frac{\Delta Y}{\Delta Z} Z_i \, ,
	\label{eq1}
\end{equation}
where $Y_0 = 0.2454$ is the primordial helium abundance and $\frac{\Delta Y}{\Delta Z} = 1.5$ is the helium enrichment ratio \citep{2020A&A...641A...6P}. 

For each of the 11 metallicities considered, we evolved a 1\,M$_\odot$ stellar model from the zero-age main sequence (ZAMS) to the terminal-age main sequence (TAMS) using identical input physics. We stored stellar structure profiles at regular intervals during the evolution, using the MESA output settings to save a profile every five timesteps.
Because the timestep size is determined adaptively by the code and the main-sequence lifetime depends on metallicity, the total number of stored profiles differs slightly from one metallicity to another. In total, the grid comprises 388 stellar models corresponding to different evolutionary stages along the main-sequence tracks.

For input physics, our models adopt the solar abundance mixture of \citet{1998SSRv...85..161G}. We took radiative opacities from the OPAL tables \citep{Iglesias1993, 1996ApJ...464..943I} and complemented them at low temperatures with the opacities of \citet{2005ApJ...623..585F}. The MESA code uses a composite equation of state that combines several formalisms \citep{Saumon1995, Timmes2000, Rogers2002, Irwin2004, Potekhin2010, Jermyn2021}. We adopted the nuclear reaction rates from JINA REACLIB \citep{Cyburt2010} and included screening effects following the prescription of \citet{Chugunov2007}.

All models include atomic diffusion, following the formulation of \citet{1994ApJ...421..828T}, due to its well-established importance in element transport in low-mass stars \citep[e.g.,][]{2018MNRAS.477.5052N}. We modeled convection with the time-dependent formalism of \citet{1986A&A...160..116K}, using a mixing-length parameter of $\alpha = 2.0$. We defined the outer boundary conditions using a gray Eddington atmosphere.

All models computed in this work do not include rotation. The connection between the thermodynamic properties of stellar interiors and observable stellar rotation remains largely unexplored. Therefore, as a first step toward understanding these relationships, we deliberately adopted nonrotating models to keep the input physics as simple and homogeneous as possible. This approach establishes a clear baseline before introducing the additional complexities associated with rotational physics.

\begin{figure*}
	\centering
	\includegraphics[width=6cm]{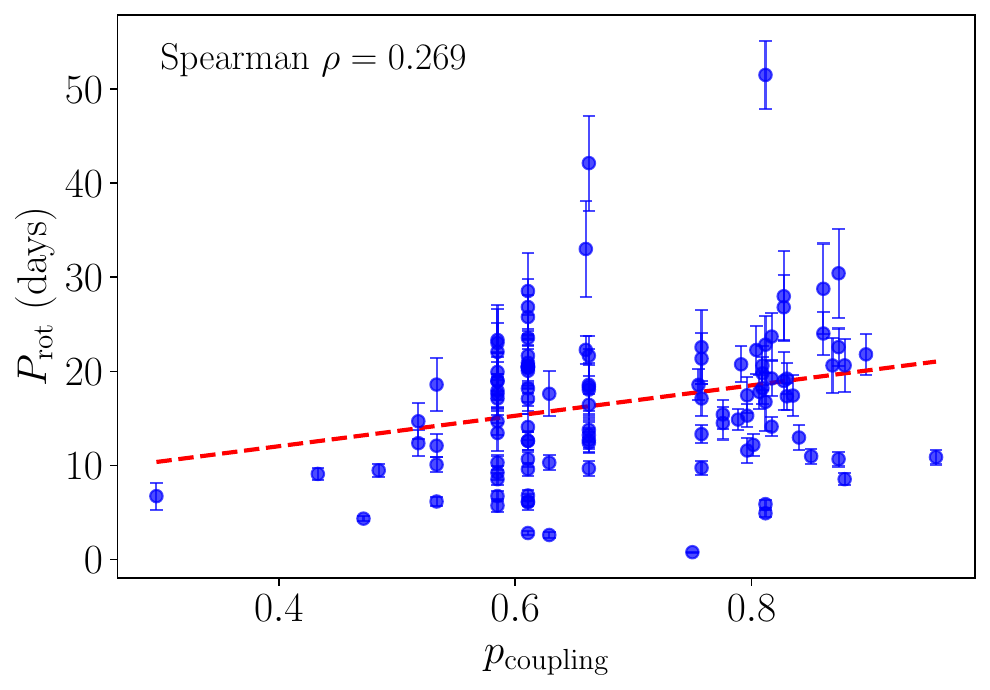}
	\includegraphics[width=6cm]{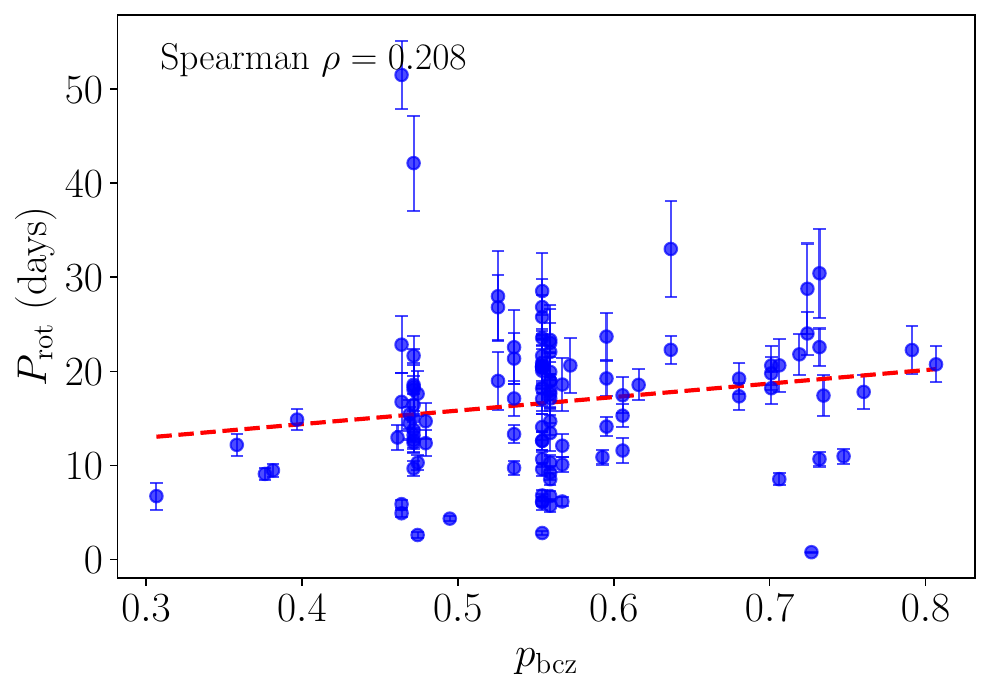}
	\includegraphics[width=5.6cm]{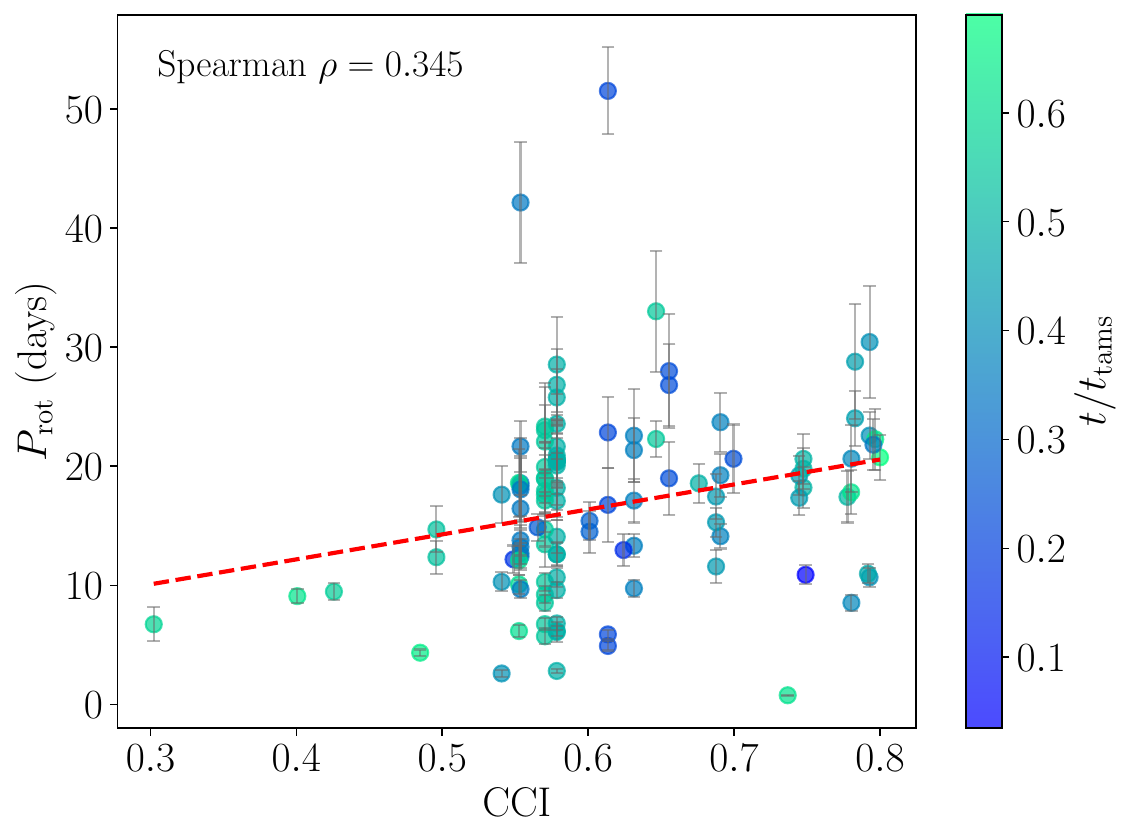}
	\caption{Dependence of stellar rotation period on fundamental convective properties for the sample of younger main-sequence stars. Left panel: Rotation period vs. percentile rank of the convective plasma parameter (\(p_{\mathrm{coupling}}\)). Middle panel: Rotation period vs. percentile rank of the convective zone depth (\(p_{\mathrm{bcz}}\)). Right panel: Rotation period vs. convective coupling index (CCI). In all panels, rotation periods are shown with error bars, and in the right panel points are color-coded by stellar age. The dashed red line indicates the linear regression fit to the data, with the associated Spearman's rank correlation coefficient (\(\rho\)) displayed.}
	\label{fig7}
\end{figure*}

\subsection{Description of the stellar sample}

From the 1404 one-solar-mass stars in the catalog described above, we aimed to identify those that could be accurately represented by our grid of theoretical models. As a first step, we excluded stars whose measured rotation periods or activity indices had uncertainties larger than the measurements themselves, resulting in a subsample of 1391 stars. Furthermore, to identify the theoretical models that best reproduce the observed stellar parameters, we implemented a $\chi^2$ minimization procedure that accounts for the uncertainties in both the effective temperature ($T_{\text{eff}}$) and the surface gravity ($\log g$). The $\chi^2$ statistic is defined as 
\begin{equation}
	\chi^2 = \frac{(T_{\text{eff, obs}} - T_{\text{eff, model}})^2}{\sigma_{T_{\text{eff}}}^2} + \frac{(\log g_{\text{obs}} - \log g_{\text{model}})^2}{\sigma_{\log g}^2 }. 
	\label{eq2}
\end{equation}
The statistic dynamically selects the appropriate positive or negative error bar ($\sigma$) for each parameter based on whether the observed value is lower or higher than the theoretical one. We then selected the theoretical point in the $T_{\text{eff}} - \log g$ plane that corresponds to the absolute minimum for a given observation as its best-match model. We retained only the results with $\chi^2 < 0.03$, as this threshold minimizes the scatter around the identity lines ($T_{\text{eff, model}}$ versus $T_{\text{eff, obs}}$ and $\log g_{\text{model}}$ versus $\log g_{\text{obs}}$). Figure \ref{fig2}  shows these identify lines and validates the robustness of our chi-square matching technique. Some scatter is always expected due to the intrinsic uncertainties in both observational and theoretical values. 

After applying this selection, we obtained a final set of 243 stars for which we could reliably extract key structural and thermodynamic properties, such as the plasma coupling parameter and the depth of the convective zone. This subsample constituted our working set, for which we established a precise mapping between the observed parameters and the underlying stellar physics. Figure \ref{fig3} shows a Kiel diagram color-coded by derived stellar ages. The nearly parallel evolutionary tracks, offset from each other, reflect the different metallicities of our model grid. The smooth monotonic gradient in normalized age ($t/t_{\mathrm{TAMS}}$), from lower fractional ages at the bottom to higher fractional ages at the top, further demonstrates that our method effectively separates the effects of metallicity from those of the evolutionary phase.

\section{Structure and internal thermodynamics: A link to rotation rates}
\label{sec:3}

In a previous study, \citet{2024A&A...690A.228B} investigated the electrostatic properties of stellar interiors using a set of models covering a range of stellar masses and evolutionary stages. They identified a connection between these electrostatic effects, quantified by the plasma coupling parameter, and the rotation rates of low-mass main-sequence stars. Their results indicate that slower rotators tend to correspond to stars with higher plasma coupling parameter values, whereas faster rotators are associated with lower values of this parameter.

In this study, we focused on stars with the same mass ($1\, M_\odot$) to examine whether the previously reported correlation between the rotation rate and the plasma coupling parameter remains valid for stars of identical mass. Furthermore, since the coupling properties of stellar interiors alone do not account for the entire diversity of observed rotation periods, we also investigated potential synergies between the structural and coupling properties that help to better understand the complex rotational behavior of main-sequence stars.

Low-mass stars, such as those considered in this work, are expected to host stellar dynamos that operate similarly to the solar dynamo \citep{2023SSRv..219...35C}. The stellar dynamo is the physical process responsible for generating and maintaining a star’s magnetic field. Rotation and the presence of convective zones are among the main ingredients of this mechanism. The very existence of convection zones arises from the microphysics and thermodynamics of stellar interiors, for instance, from high opacities that hinder radiative energy transport or from ionization regions where ionized gas efficiently absorbs photons.

In this section, we describe the key properties of our stellar models, focusing on the characteristics of their convective zones from both a structural perspective (e.g., the location of the base of the convective zone) and a thermodynamic perspective, quantified by the plasma coupling parameter. We also introduce a new index that combines structural information, specifically the thickness of the convective zones, with the corresponding values of the global plasma coupling parameter.

\subsection{The global characterization of plasma coupling in convective zones}

\begin{table*}[htbp]
	\centering
	\caption{Younger main-sequence stars}
	\label{tab1}
	\footnotesize
	\begin{tabular}{lcccc}
		\toprule
		Comparison & Spearman & Pearson & Kendall & Regression \\
		\midrule
		\text{$P_{\mathrm{rot}}$ vs $p_{\mathrm{coupling}}$} & 
		$\rho = 0.269$ & $r = 0.257$ & $\tau = 0.192$ & $R^2 = 0.066$ \\
		& (0.087--0.434) & (0.085--0.418) & (0.057--0.322) & RMSE = 7.417 \\
		& $p = 4.38 \times 10^{-3}$ & $p = 6.41 \times 10^{-3}$ & $p = 3.89 \times 10^{-3}$ & MAE = 5.554 \\
		\addlinespace
		
		\text{$P_{\mathrm{rot}}$ vs $p_{\mathrm{bcz}}$} & 
		$\rho = 0.208$ & $r = 0.175$ & $\tau = 0.148$ & $R^2 = 0.031$ \\
		& (0.025--0.386) & ($-$0.019--0.388) & (0.009--0.276) & RMSE = 7.557 \\
		& $p = 2.88 \times 10^{-2}$ & $p = 6.54 \times 10^{-2}$ & $p = 2.57 \times 10^{-2}$ & MAE = 5.490 \\
		\addlinespace
		
		\text{$P_{\mathrm{rot}}$ vs CCI} & 
		$\rho = 0.345$ & $r = 0.250$ & $\tau = 0.248$ & $R^2 = 0.062$ \\
		& (0.177--0.501) & (0.099--0.409) & (0.122--0.363) & RMSE = 7.432 \\
		& $p = 2.08 \times 10^{-4}$ & $p = 8.15 \times 10^{-3}$ & $p = 1.92 \times 10^{-4}$ & MAE = 5.406 \\
		\bottomrule
	\end{tabular}
	\tablefoot{
		 Correlation coefficients (Spearman's $\rho$, Pearson's $r$, and Kendall's $\tau$) and linear regression metrics for the relationship between stellar rotation period ($P_{\mathrm{rot}}$) and the percentile ranks for the depth of the convective zone and for the convective plasma parameter. Confidence intervals and p-values are provided for correlation coefficients.
	}
\end{table*}

The plasma coupling parameter, $\Gamma$, is a dimensionless quantity defined as the ratio between the average Coulomb potential energy of interaction among particles and their average thermal kinetic energy. We considered the ionic plasma coupling parameter $\Gamma_i$, given by  

\begin{equation}
	\Gamma_i = \frac{Z_i^2 e^2}{a_i k T},
	\label{eq3}
\end{equation}
where $Z_i$ is the mean ionic charge, $T$ is the local temperature, $e$ is the elementary charge, $k$ is the Boltzmann constant, and $a_i$ is the Wigner--Seitz radius, which represents the mean inter-ionic distance \citep[e.g.,][]{1986ApJS...61..177P}.  

To characterize the overall degree of plasma coupling within the convective zones of our models, we adopted the global plasma coupling parameter, introduced by \citet{2024A&A...690A.228B}, defined as  

\begin{equation}
	\overline{\Gamma}_{\mathrm{conv}}= \frac{\int_{M_{\mathrm{bcz}}}^{M_{\mathrm{surf}}} \Gamma_i \, dm(r)}{M_{\mathrm{conv}}},
	\label{eq4}
\end{equation}
where 
$dm(r) = 4 \pi r^2 \rho \, dr$, $M_{\mathrm{bcz}}$ and $M_{\mathrm{surf}}$ denote the mass coordinates at the base and surface of the convective zone, respectively, and $M_{\mathrm{conv}}$ is the total mass of the convective outer envelope.  

A higher value of $\overline{\Gamma}_{\mathrm{conv}}$ corresponds to a stronger degree of plasma coupling within the convective zone of a model. Figure \ref{fig4} shows the variation of the global convective plasma parameter along the main sequence for stellar models of different metallicities. The figure highlights the clear influence of metallicity on the degree of plasma coupling in the convective zones: higher metallicity leads to higher values of $\overline{\Gamma}_{\mathrm{conv}}$, indicating that collective plasma effects become more significant in metal-rich stars.

\subsection{Depth of the convective zones}

Within the stellar dynamo mechanism, convective motions in the outer envelope and stellar rotation act together to amplify and maintain an organized magnetic field. For stars with masses in the range $0.35 \, M_\odot \lesssim M \lesssim 1.3\,M_\odot$, the thickness of the convective zone plays a crucial role in dynamo action for several reasons. 

First, the convective zone constitutes the turbulent region where the dynamo operates. Second, a deeper convective zone can be associated with a more powerful dynamo engine; for instance, in the context of the $\alpha$-effect, helical turbulence converts toroidal magnetic fields into poloidal components. Moreover, a greater convective depth implies a broader range of pressure and density conditions, allowing magnetic flux tubes to rise more efficiently due to buoyancy. 

Finally, and perhaps most importantly, a thicker convective zone contributes to the stability of the tachocline, the stratified shear layer at the interface between the radiative interior and the convective envelope. The tachocline plays a fundamental role in stretching the poloidal field into a toroidal one through differential rotation (the so-called $\Omega$-effect). Therefore, the thickness of the outer convective zone is a key parameter in determining the possible configurations and efficiency of the stellar dynamo.  

For $1\,M_\odot$ stellar models, one typically recalls the solar value of approximately $0.3\,R_\odot$, which corresponds to the depth of the solar convective zone. This value, however, applies specifically to a star with solar metallicity and solar age. In our set of models, both the metallicity and the age vary significantly, leading to a wide diversity in the depths of the convective zone bases. 

Figure \ref{fig5} illustrates this diversity for our $1\,M_\odot$ models, showing that the base of the convective zone can lie anywhere from less than 1\% below the stellar surface to as deep as 38\% of the stellar radius. Understanding this variation in convective zone extent is essential for predicting and modeling stellar magnetic activity, stellar winds, and the observed rotation rates of low-mass stars.

\subsection{The convective coupling index}

\begin{figure*}
	\centering
	\includegraphics[width=6cm]{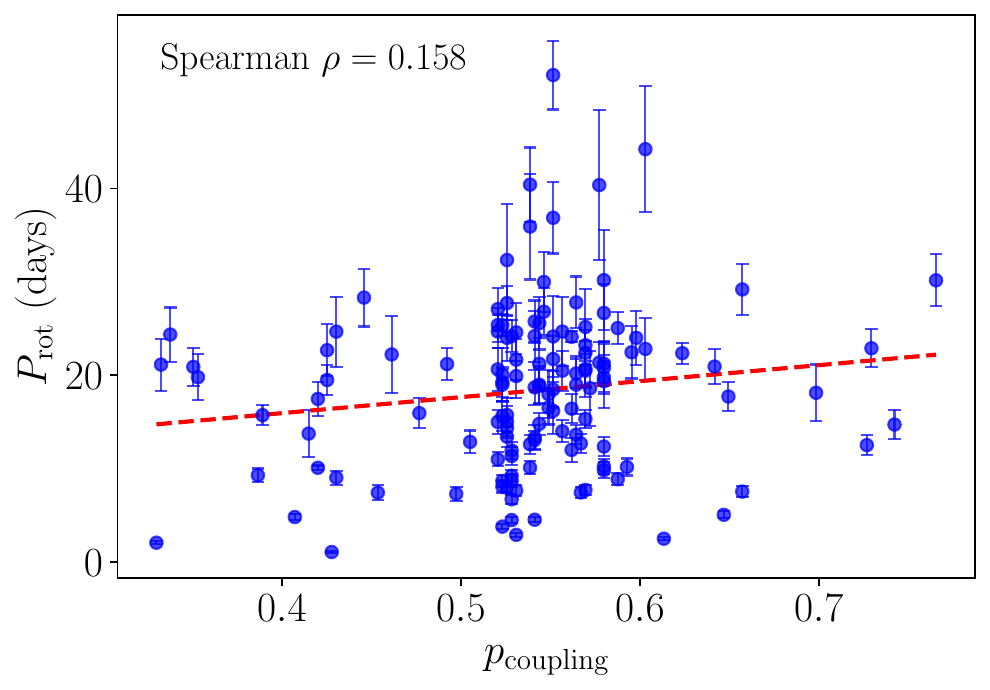}
	\includegraphics[width=6cm]{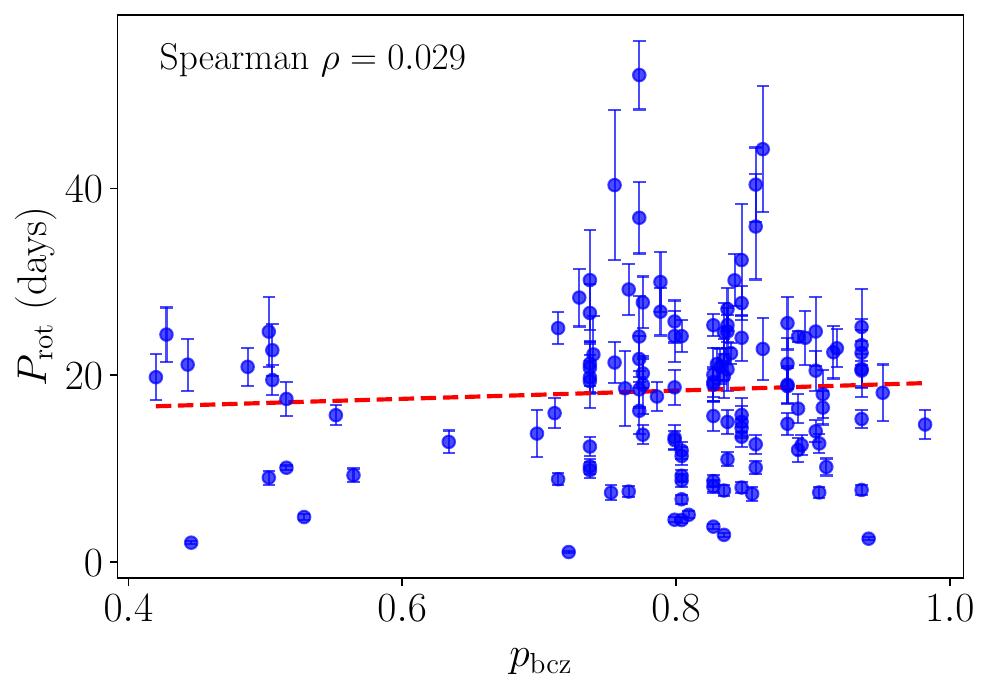}
	\includegraphics[width=5.6cm]{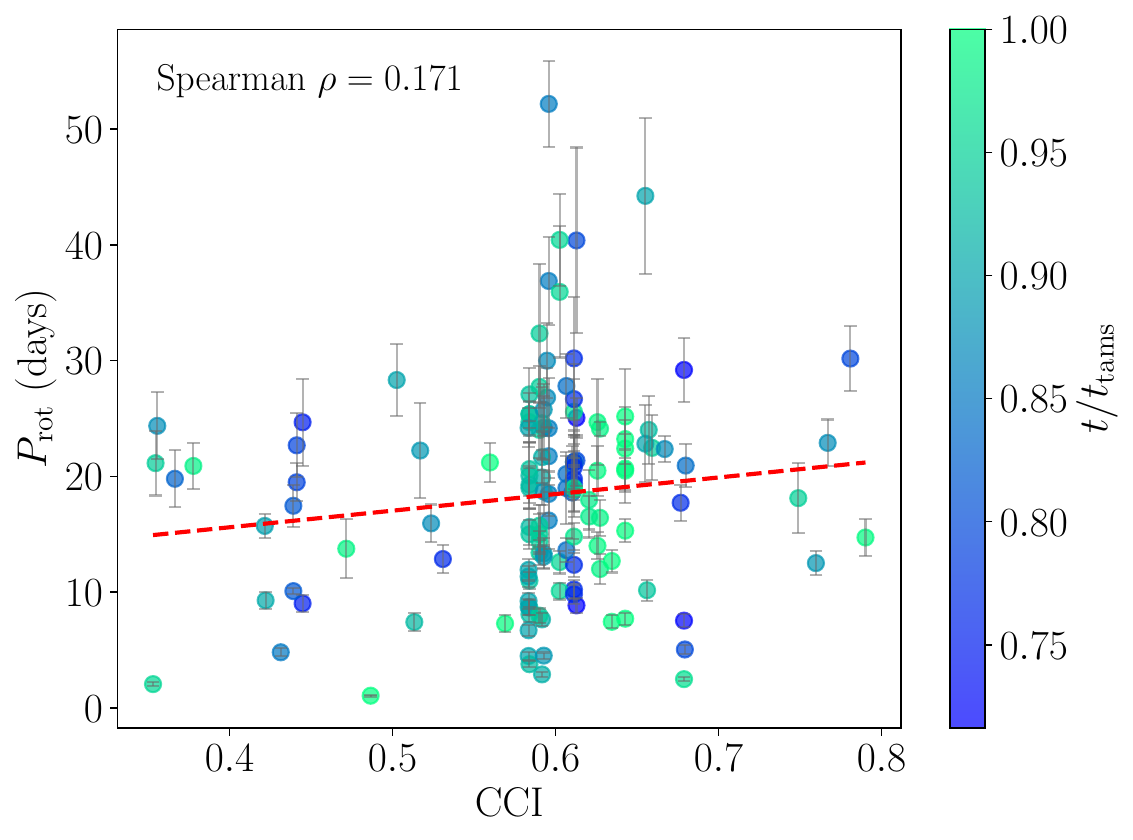}
	\caption{Panels and symbols follow the same conventions as in Figure \ref{fig7}, but for the sample of older main-sequence stars.}
	\label{fig8}
\end{figure*}

Figure~\ref{fig6} shows that the linear trend of the convective plasma parameter with effective temperature (Spearman $\rho = -0.27$, p-value $= 1.4\times10^{-5}$) exhibits a similar behavior to that of the rotation periods with effective temperature (Spearman $\rho = -0.21$, p-value $= 1.2\times10^{-3}$). For reference, Figure \ref{fig6} includes the observed solar values $T_{\mathrm{eff},\odot} = 5772\,\mathrm{K}$ and $P_{\mathrm{rot},\odot} = 27.3\,\mathrm{days}$ \citep{2016AJ....152...41P, 2000SoPh..191...47B}. The corresponding theoretical quantities derived from the MESA standard solar model (computed with the GS98 metallicity) are $T_{\mathrm{eff},\odot} = 5767\,\mathrm{K}$ and $\overline{\Gamma}_{\mathrm{conv},\odot} = 0.089$.
		
This result is noteworthy because it reproduces the main conclusion of \citet{2024A&A...690A.228B}, although in a different context. 
\citet{2024A&A...690A.228B} explored the connection between electrostatic effects and stellar rotation rates using a grid of models that span a range of stellar masses ($0.7\,M_\odot$ to $1.4\,M_\odot$), ages, and metallicities. 
In the present study, we find a similar relationship between the convective plasma parameter, which characterizes the degree of plasma coupling, and the rotation rates of our sample of stars. 
This shows that the correlation between the electrostatic properties of the plasma and stellar rotation rates can also be recovered among stars of the same mass, in this case $1\,M_\odot$.  

However, these correlations are weak and do not fully account for the observed diversity in stellar rotation rates, as rotation is a complex physical process influenced by multiple interrelated factors.

We next investigated the direct correlations between stellar rotation rates (measured via rotation periods), the convective plasma parameter, and the extent of the convective zones, treating each separately. 
We introduced an index that combines both structural information, represented by the extent of the convective zones, and thermodynamic information, characterized by the global convective plasma parameter ($\overline{\Gamma}_{\mathrm{conv}}$) defined above. To construct a meaningful composite index, we first normalized the two physical quantities to a common, dimensionless scale before combining them. 
For this purpose, we applied a percentile-based normalization as follows:  
(1) we computed the percentile rank of each value of the location of the base of the convective zone (BCZ) within the full range of BCZ values in our sample, and  
(2) we compute the percentile rank of each value of $\overline{\Gamma}_{\mathrm{conv}}$ within the full range of $\overline{\Gamma}_{\mathrm{conv}}$ values in the sample.  

We then combined both percentiles into a single index, which we refer to as the convective coupling index (CCI), defined as  

\begin{equation}
	\mathrm{CCI} = w_1 \times p_{\mathrm{bcz}} + w_2 \times p_{\mathrm{coupling}} ,
	\label{eq5}
\end{equation}
where $p_{\mathrm{bcz}}$ and $p_{\mathrm{coupling}}$ represent the percentile ranks corresponding to the location of the BCZ and the values of $\overline{\Gamma}_{\mathrm{conv}}$, respectively. 
The coefficients $w_1$ and $w_2$ are the weighting factors that we discuss in the next section.  

This dimensionless quantity, which ranges from 0 to 1, captures the combined relative influence of two key stellar properties: the depth of the convective envelope and the degree of plasma coupling within the convective zone.  

We expect stars with deeper convective zones and stronger Coulomb coupling to exhibit longer rotation periods, as these conditions favor more efficient magnetic braking. 
Consequently, we anticipate a positive correlation between the CCI and the rotation period.

\begin{table*}[htbp]
	\centering
	\caption{Older main-sequence stars}
	\label{tab2}
	\footnotesize
	\begin{tabular}{lcccc}
		\toprule
		Comparison & Spearman & Pearson & Kendall & Regression \\
		\midrule
		\text{$P_{\mathrm{rot}}$ vs $p_{\mathrm{coupling}}$} & 
		$\rho = 0.158$ & $r = 0.145$ & $\tau = 0.106$ & $R^2 = 0.021$ \\
		& ($-$0.006--0.313) & ($-$0.017--0.301) & ($-$0.001--0.217) & RMSE = 8.679 \\
		& $p = 7.02 \times 10^{-2}$ & $p = 9.80 \times 10^{-2}$ & $p = 7.38 \times 10^{-2}$ & MAE = 6.793 \\
		\addlinespace
		
		\text{$P_{\mathrm{rot}}$ vs $p_{\mathrm{bcz}}$} & 
		$\rho = 0.029$ & $r = 0.061$ & $\tau = 0.018$ & $R^2 = 0.004$ \\
		& ($-$0.131--0.194) & ($-$0.082--0.214) & ($-$0.092--0.124) & RMSE = 8.755 \\
		& $p = 7.40 \times 10^{-1}$ & $p = 4.89 \times 10^{-1}$ & $p = 7.59 \times 10^{-1}$ & MAE = 6.846 \\
		\addlinespace
		
		\text{$P_{\mathrm{rot}}$ vs CCI} & 
		$\rho = 0.171$ & $r = 0.128$ & $\tau = 0.114$ & $R^2 = 0.016$ \\
		& ($-$0.003--0.331) & ($-$0.017--0.286) & (0.000--0.224) & RMSE = 8.699 \\
		& $p = 4.96 \times 10^{-2}$ & $p = 1.434 \times 10^{-1}$ & $p = 5.51 \times 10^{-2}$ & MAE = 6.816 \\
		\bottomrule
	\end{tabular}
	\tablefoot{
		Correlation coefficients (Spearman's $\rho$, Pearson's $r$, Kendall's $\tau$) and linear regression metrics for the relationship between stellar rotation period ($P_{\mathrm{rot}}$) and the percentile ranks for the depth of the convective zone and for the convective plasma parameter. Confidence intervals and p-values are provided for correlation coefficients.
	}
\end{table*}

\section{Enhanced rotation-CCI correlation in younger main-sequence stars}\label{sec:4}

Figure~\ref{fig4} shows that the behavior of $\overline{\Gamma}_{\mathrm{conv}}$ differs between the first $\sim70\%$ of the main-sequence lifetime and the remaining $\sim30\%$. 
The values of $\overline{\Gamma}_{\mathrm{conv}}$ decrease more steeply during the early main-sequence phase for almost all metallicities. 
In the later stages, the decrease in $\overline{\Gamma}_{\mathrm{conv}}$ slows down and may even reverse near the terminal-age main sequence (TAMS). 
The exception occurs for stellar models with the lowest metallicities ($Z = 0.0005$ and $Z = 0.001$), for which $\overline{\Gamma}_{\mathrm{conv}}$ continues to decrease throughout the main sequence.  

This general behavior of the global plasma parameter is consistent with the results obtained when the parameter is computed over the entire stellar structure \citep{2024A&A...690A.228B}, rather than restricted to the convective zones, as in the present study. 
From a physical perspective, this indicates that the degree of plasma coupling decreases as a star evolves along the main sequence, reaching a minimum around $70\%$ of its main-sequence lifetime.

To investigate possible correlations between $\overline{\Gamma}_{\mathrm{conv}}$ and the observed rotation rates, we divided the analysis into two parts, motivated by the behavior of $\overline{\Gamma}_{\mathrm{conv}}$ described above. 
We separated the full sample of 243 stars into two subsamples according to stellar age. 
The first subsample includes stars located within the first $70\%$ of their main-sequence lifetime, whereas the second subsample comprises stars in the final $30\%$ of the main-sequence evolution. 
We refer to these groups as younger main-sequence and older main-sequence stars, respectively. 
Using these two subsamples, we analyzed the correlations between the two percentile ranks (corresponding to the BCZ depths and $\overline{\Gamma}_{\mathrm{conv}}$) and the rotation rates separately. 
We then investigated whether the correlation improves when the CCI is considered, which accounts for the combined influence of both structural and thermodynamic properties, namely, the depth of the convective zones and the plasma coupling characteristics within them.

\subsection{Statistical calibration of the convective plasma parameter index}

Our analysis used multiple statistical approaches to characterize the relationships described above. We computed different correlation coefficients (Spearman, Pearson, and Kendall) together with bootstrapped confidence intervals. Specifically, we obtain 95\% confidence intervals through bootstrapping with 1000 resamples to ensure robust uncertainty estimates.

For the CCI, we implemented an optimization procedure to determine the optimal weighting between the two percentiles ($p_{\mathrm{bcz}}$ and $p_{\mathrm{coupling}}$), with the goal of maximizing the correlation with the stellar rotation periods. 
The optimization follows a grid-search approach, testing 100 different weight combinations ($w_1$ ranging from 0 to 1 in steps of 0.01, with $w_2 = 1 - w_1$). 
For each combination of weights, we calculated the three correlation coefficients (Spearman, Pearson, and Kendall) as well as regression metrics ($R^2$, RMSE, and MAE).  

We selected the optimal combination of weights by maximizing the Spearman correlation coefficient, which we adopted as the primary criterion because of its robustness against nonlinear relationships. 
Furthermore, we validated the optimized CCI through statistical comparison tests to assess whether it significantly outperformed the individual predictors ($p_{\mathrm{bcz}}$ and $p_{\mathrm{coupling}}$). 
We also applied a five-fold cross-validation to evaluate the stability of the optimized weights across different data subsets.  
Overall, this optimization framework ensures that the resulting CCI weighting is both statistically optimal and robust against overfitting, providing a data-driven combination of the two physical parameters for investigating and predicting stellar rotation behavior.

\subsection{Younger main-sequence stars}

Figure~\ref{fig7} shows the correlations between $p_{\mathrm{bcz}}$, $p_{\mathrm{coupling}}$, and the CCI index with the rotation period for the younger main-sequence stars and  Table~\ref{tab1} provides a summary of the statistical results.
When considered individually, both $p_{\mathrm{bcz}}$ and $p_{\mathrm{coupling}}$ exhibit statistically significant but modest positive correlations with the stellar rotation period. 
The combined CCI index (Eq.~\ref{eq5}) yields a notable improvement, with a Spearman coefficient of $\rho = 0.345$, corresponding to a 28\% increase relative to $p_{\mathrm{coupling}}$ alone and a 66\% increase relative to $p_{\mathrm{bcz}}$ alone. 
The nearly equal optimal weighting--57\% for the influence of the convective-zone depth ($w_1$) and 43\% for the coupling properties of the plasma ($w_2$)--indicates that both quantities contribute substantially to predicting stellar rotation periods.

This synergy between the depth of the convective zone and the plasma coupling degree clearly enhances the correlation with rotation rates, suggesting that these two mechanisms jointly influence angular-momentum evolution. 
In this sample of younger main-sequence stars, the percentile rank of the convective plasma parameter emerges as a physically important contributor. 
Although the correlation with the CCI remains moderate and the proportion of variance explained ($R^2 = 0.066$ for $p_{\mathrm{coupling}}$ alone and $R^2=0.062$ for the CCI may appear small, these values still represent statistically significant and physically meaningful trends in the context of stellar rotation dynamics.
Stellar rotation is a function of multiple interconnected processes, including magnetic braking efficiency, internal angular momentum transport, stellar winds and mass loss, initial rotational conditions, different magnetic topologies, and metallicity and age uncertainties. In this context, isolating the contribution of only two parameters (one structural and one thermodynamic) is not expected to account for a large fraction of the total variance in the data.
Moreover, the existence of observational uncertainties in rotation periods and atmospheric parameters naturally reduces the fraction of variance in the rotation period that can be explained by a simple linear model.
It is also important to note that we evaluated the correlations through the Spearman coefficient, which is sensitive to monotonic trends even when the relationship is not strictly linear. In such cases, the R$^2$ derived from a linear fit may underestimate the strength of the underlying monotonic association.
Therefore, even with the modest R$^2$ values, the statistically significant correlation coefficients indicate the presence of a genuine, although partial, physical connection between convective properties and stellar rotation in the younger main-sequence regime.

\subsection{Older main-sequence stars: The fading signature}

For the sample consisting of older main-sequence stars, we do not obtain an enhanced correlation with the rotation periods when considering the combined influence of the convective-zone depth and the plasma coupling degree. 
Figure~\ref{fig8} shows the correlations between $p_{\mathrm{bcz}}$, $p_{\mathrm{coupling}}$, the CCI index, and the rotation rates.
We find that none of the correlations are statistically significant, in contrast to the results for the younger stars. 
Interestingly, the percentile rank $p_{\mathrm{coupling}}$ becomes relatively more influential.  

The optimization process yields a different balance between $w_1$ and $w_2$: 
the convective-zone depth contributes only 20\% ($w_1$), whereas the coupling properties of the plasma contribute 80\% ($w_2$). 
This pattern suggests that as stars evolve along the main sequence, the relative influence of the convective-zone depth on rotation weakens ($\rho = 0.029$, $p = 0.074$), 
while the thermodynamic properties associated with the plasma coupling retain some influence ($\rho = 0.158$, $p = 0.07$), 
becoming almost the sole contributor to the weakened CCI correlation (see Table \ref{tab2}).

\section{Conclusions}
\label{sec:5}

In this work, we investigate whether the electrostatic properties of stellar convective zones, together with their structural characteristics, specifically the depth of the convective zones, can be linked to the observed rotation rates of a sample of $1\,M_\odot$ stars. 
To this end, we introduced an index that combines structural information, expressed through the percentile ranks of the convective-zone depths, with the percentile ranks of the plasma coupling parameter integrated over the stellar convective zones. 
We then studied the correlations between each percentile rank individually and between their combined effect using the CCI index introduced in Section~\ref{sec:3}. 
Our main conclusions are as follows.

\begin{enumerate}
	\item Younger main-sequence stars (those with ages less than 70\% of their main-sequence lifetime) exhibit a statistically significant correlation between the rotation period and the combined structural and thermodynamic properties of their convective zones ($\rho = 0.345$, $p = 2.08 \times 10^{-4}$). 
	In a context characterized by high complexity and multiple interconnected physical processes, these statistical results likely indicate a genuine underlying physical relationship.
	
	These findings also suggest that the coupling properties of plasma in the convective zones of low-mass, main-sequence stars may be linked to the observational phenomenon known as weakened magnetic braking \citep{2024ApJ...962..138S, 2024MNRAS.533.1290S}. 
 \citep{2024A&A...690A.228B} propose a connection between plasma coupling and weakened magnetic braking using a grid of stellar models that span different masses, metallicities, and ages.
	The present work finds a similar relation using a grid of models all having the same stellar mass, providing independent verification of the earlier result.
	
	\item For the older main-sequence stars (those older than 70\% of their main-sequence lifetime), the correlations between the rotation periods and the CCI index weaken considerably and lose statistical significance. 
	However, this evolutionary change is scientifically meaningful, as it reflects a shift in the physical processes governing angular-momentum evolution. 
	Interestingly, while the overall correlations decrease with age, the plasma coupling properties become relatively more influential, accounting for 80\% of the CCI weighting.
\end{enumerate}

The present work can be extended in several ways. In particular, applying the same framework to stars of different masses would allow us to explore how the interplay between convective structure and plasma coupling evolves across distinct rotational regimes, for example, below and above the Kraft break (e.g., around $0.8\,M_\odot$ and $1.3\,M_\odot$). Such an analysis would provide valuable insight into how the mass-dependent internal structure influences the evolution of the angular-momentum. Extending this approach further to lower masses, i.e., to partially convective M dwarfs, is especially interesting, given the importance that electrostatic interactions assume in their interiors. This will become feasible with the upcoming PLATO (PLAnetary Transits and Oscillations of stars) mission \citep{2025ExA....59...26R}. Incorporating rotating stellar models in future studies could also enable a more self-consistent treatment of angular-momentum transport and its feedback on internal structure within this context.

Finally, we place this study in a broader astrophysical context. 
For low-mass main-sequence stars, rotation and magnetic activity are tightly linked through the well-established rotation--activity relationship \citep[e.g.,][]{1984ApJ...279..763N, 2025A&A...702A.218F}. 
Magnetic activity, in turn, arises from an underlying dynamo mechanism operating in stellar interiors, particularly within their convective zones. 
Stellar dynamo processes depend sensitively on transport coefficients such as viscosity and magnetic diffusivity (or electrical conductivity). 
The equation of state, which sets temperature and density profiles, determines the plasma coupling parameter, which in turn controls the physical processes that govern the transport coefficients used in stellar dynamo theory.

\begin{acknowledgements}
	The authors thank the referee for the valuable comments and suggestions that have helped improve the manuscript. This work was supported by the Fundação para a Ciência e Tecnologia (FCT), Portugal, through grants UID/PRR/00099/2025 (\href{https://doi.org/10.54499/UID/PRR/00099/2025}{10.54499/UID/PRR/00099/2025}) and UID/00099/2025 (\href{https://doi.org/10.54499/UID/00099/2025}{10.54499/UID/00099/2025}), attributed to the Center for Astrophysics and Gravitation (CENTRA/IST/ULisboa).
\end{acknowledgements}


\bibliographystyle{aa} % style aa.bst
\bibliography{bib_file} % your references Yourfile.bib

\end{document}